\documentclass[preprint,nofootinbib,showpacs,aps]{revtex4}
\def\be{\begin{equation}}
\def\ee{\end{equation}}
\begin{document}
\title{Thermodynamic scheme of inhomogeneous perfect fluid mixtures}
\author{Rub\'en D. Z\'arate}
\email{zarate@nuclecu.unam.mx}
\affiliation{ Instituto de Ciencias Nucleares\\
     Universidad Nacional Aut\'onoma de M\'exico \\
     P.O. Box 70-543, M\'exico D.F. 04510}

\author{Hernando Quevedo}\email{quevedo@physics.ucdavis.edu}
\affiliation{Instituto de Ciencias Nucleares\\
Universidad Nacional Aut\'onoma de M\'exico\\
 P.O. Box 70-543, M\'exico D.F. 04510\\}
\affiliation{Department of Physics\\ University of California \\
Davis, CA 95616}

\begin{abstract}
We analyze the compatibility between the geometrodynamics and
thermodynamics of a binary mixture of perfect fluids which
describe inhomogeneous cosmological models. We generalize the
thermodynamic scheme of general relativity to include the
chemical potential of the fluid mixture with non-vanishing
entropy production. This formalism is then applied to the case of
Szekeres and Stephani families of cosmological models. The
compatibility conditions turn out to impose symmetry conditions
on the cosmological models in such a way that only the limiting
case of the Friedmann-Robertson-Walker model remains compatible.
This result is an additional indication of the incompatibility
between thermodynamics and relativity.
\end{abstract}

 \noindent \noindent % 03.70.+k, 04.62.+vhttp://www.iop.org/journals/submit

\maketitle

\section{Introduction}
When considering cosmological questions, it is a conventional wisdom
to work with the homogeneous and isotropic Friedmann-Robertson-Walker (FRW)
 models,
since they reproduce many of the features of the observable
universe. Nevertheless, the desire to describe details such as
the process of structure formation, which leads to local
inhomogeneities, requires the use of more general models. A
natural framework to do this is found in the inhomogeneous
cosmological models, which are exact solutions to Einstein's field
equations, representing inhomogeneous universes and which contain
the FRW models in a certain geometric limit, thus, generalizing
them. Although the presence of inhomogeneities is crucial in the
process of structure formation, especially in the context of
galaxy formation, the inhomogeneous models to be 
investigated in this work would be insufficient to take into
account all the details of the physics of structure formation 
(see, for instance,
\cite{chou} for a recent review).

The most general classes of inhomogeneous models  known are
the Stephani \cite{ste} and Szekeres \cite{sze}solutions, which
are irrotational perfect fluid exact solutions and have, in
general, no isometries. Both families admit a barotropic equation
of state only in the FRW limit, and this has risen the question
of how to find or generate non trivial but physically meaningful
equations of state. To this question is attached the fact that an
evolving universe is, from the thermodynamics point of view, a
system out of equilibrium, thus, we must make sure that defining
thermodynamic variables makes sense. In this direction, Coll and
Ferrando \cite{colfer} have shown that an exact solution admits a
``thermodynamic scheme", that is, has well defined thermodynamic
variables compatible with the field equations, provided the
integrability conditions of Gibbs equation are satisfied, what
turns out to be equivalent to the existence of equations of
state, not necessarily barotropic.

Following the thermodynamic scheme proposed in \cite{colfer},
 Krasi\'nski, Quevedo and Sussman \cite{ahr,hr1,hr2} worked
out the thermodynamic interpretation of the inhomogeneous cosmological models,
and found that, in the
general case, defining sound thermodynamics involves forcing isometries on the
metrics. However, such analysis was made for a one component perfect fluid,
complying with particle conservation, which satisfies automatically
$\dot{s} = 0$, {\it i.e.}, for a flow with null entropy production.

In an attempt to understand such incompatibility between thermodynamics and
perfect fluid inhomogeneous cosmological models, in this work
we study the simplest
non dissipative irreversible case in
which the entropy production is in general non vanishing: a binary mixture of
perfect fluids. To this end, we generalize the thermodynamic scheme to the case
of a binary mixture, and analyze both the integrability conditions of Gibbs
equation and the condition of entropy production, due to the fluids mixing
process. From the integrability conditions we are able to recognize the
equations of state which describe the mixture, and which do not involve imposing
isometries on the metrics. However, to satisfy the entropy production condition
the metrics must reduce to those of the FRW cosmologies, a result that we
interpret as a further indication to the incompatibility between
thermodynamics and relativity.

Different aspects of binary mixtures of perfect fluids have been analyzed
in general relativity. Two-fluid models have been intensively applied to
describe cosmological models in which one fluid represents radiation and
the second one the matter content of the universe
\cite{radmat1,radmat2,radmat3,radmat4,radmat5,
radmat6,radmat7,radmat8,radmat9}. If the
two-fluids are comoving, then their energy-momentum tensor is
effectively that of a single fluid. If the fluids are not comoving,
then Letelier \cite{let1} has shown that the energy-momentum tensor
can be transformed into a tensor explicitly exhibiting a preferred
spatial direction which resembles the energy-momentum tensor of a
viscous fluid. It was also shown in \cite{let1} and \cite{let2}
that an anisotropic fluid
can be consistently described by two-perfect-fluid components.
Krisch and Smalley \cite{krisma}
have investigated the propagation of discontinuities in
the relativistic two-fluid system  described by Letelier's
tensor. Zimdahl, Pavon and Maartens \cite{zim1,zim2}
have studied inflationary models as a mixture of two
interacting and reacting fluids within the framework of
irreversible thermodynamics. The authors consider an out of
equilibrium dissipative mixture, with irreversible evolution
in order to describe the reheating process in inflationary universe
models. For this end, they assume the condition $\dot{s}_{A} = 0$ for
each component $A$ of the mixture. Finally, Cissoko \cite{cis}
has analyzed a mixture of a reacting and coupled two perfect fluids
flowing with distinct 4-velocities and introduced the Lorentz factor,
associated with the relative velocity of the fluids, as an additional
thermodynamic variable. In all these works less attention has been paid
to the study of the compatibility between the thermodynamic and
geometric evolution of the models, specially when inhomogeneities are
present.

This paper is structured as follows. In Sections II and III we study
the thermodynamics of a binary mixture of comoving perfect fluids,
derive the expression for the entropy production due to the mixing
process, and propose a thermodynamic scheme which explicitly includes
the chemical potential. In Section IV we apply the conditions for
the existence of a thermodynamic scheme to the Stephani and Szekeres
inhomogeneous cosmological models. Finally, in Section V we discuss our
results and comment on additional problems for further research.

\section{Thermodynamics of the source}

Cosmology addresses, among other questions, the one concerning the process of
evolution of the universe as a whole. Such process, according to general
relativity, involves the evolution of both the source and the geometry,
connected through Einstein's field equations. Besides satisfying the field
equations, the source must be subject to the laws of thermodynamics, which
formally, involve quantities defined only in equilibrium. Thus, to describe an
evolving system, from the thermodynamic viewpoint, we must extend the concepts
and laws to the non equilibrium case. The simplest way to do this, and the most
invoked one in cosmology (implicitly or explicitly), is based on the local
equilibrium hypothesis, which asserts that every point of a system not too far
from equilibrium, has a neighborhood which is in equilibrium, that is, in such a
neighborhood we can define uniquely all thermodynamic variables, with the same
physical meanings as in equilibrium and satisfying all the relations the
variables satisfy in equilibrium. Thus, the gradients and temporal changes of
the variables in the system are connected through the relations they satisfy in
equilibrium,{\it i.e.}, we must treat them as scalar fields connected through the
equilibrium relations. With this in mind, we now consider the source of the
field in our models: a perfect fluid consisting of a binary mixture of perfect
fluids. Its energy-momentum tensor is

\begin{equation}
T_{\alpha\beta} = (\rho+p)u_\alpha u_\beta + pg_{\alpha\beta}
\label{1}
\end{equation}
where $\rho$ and $p$ are the total matter-energy density and the
mixture's pressure, respectively, and $u_\alpha, \
(\alpha=0,1,2,3)$ is the $4$-velocity associated with the total
matter flow. Such energy-momentum tensor corresponds to a
continuum in which energy transport occurs only due to matter
transport. Thus, entropy production will be a consequence of the
mixing process alone, as the composition of the mixture will be,
in general, inhomogeneous. This tensor satisfies the conservation
law $T^{\alpha\beta}_{\ \ ;\beta} = 0$, which in particular,
 implies the contracted Bianchi identity
\begin{equation}
\dot\rho + (\rho +p)\Theta = 0
\label{2}
\end{equation}
where $ \Theta = u^\alpha_{\ ;\alpha}$ is the scalar expansion.
Besides, we restrict ourselves to the case in which particles of
both components are conserved
\begin{equation}
(nu^\alpha)_{;\alpha} = 0
\label{3}
\end{equation}
where $n$ is the total particle number density. The analogues of equations
(\ref{2}) and (\ref{3}) for the case of  a one component perfect fluid,
together with the corresponding Gibbs equation imply, automatically, null
entropy production, $\dot s = 0$.
However, for the case of a binary mixture, Gibbs equation reads
\begin{equation}
Tds = d(\rho/n) + pd(1/n) - \mu_1d c - \mu_{2}d c_{2}
\end{equation}
where $s$ is the entropy per particle, $T$ is the temperature, $\mu_{1}$ and
$\mu_{2}$ are the chemical potentials of components $1$ and $2$, and $c$ and
$c_{2}$ are the fractional concentrations ($c + c_{2} = 1$) of each kind of
particle in the fluid element under consideration. It is clear that the fluid
composition is determined by only one parameter, so we can write
\begin{equation}
Tds = d(\rho/n) + pd(1/n) - \mu d c
\label{5}
\end{equation}
where we have introduced the mixture's ``chemical potential",
$ \mu = \mu_{1} - \mu_{2}$.
Equation (\ref{5}),  together with (\ref{2}) and (\ref{3}), now implies
\begin{equation}
{\dot s} = -{\mu\over T}{\dot c}
\label{6}
\end{equation}
This corresponds to the entropy production in a fluid in which the mixing
process is the only irreversible process taking place, and it depends only on
the change of the Gibbs free energy, due to change in composition, and on
temperature. This represents the main difference with respect to
the case of a one component fluid, and it is our interest to exhibit its
consequences.

\section{Gibbs $1$-form}

As equation (\ref{6}) states, to describe the thermodynamics of any fluid
element of the mixture, we must introduce the conjugate pair of variables $\mu$
and $c$, in terms of which the entropy production is written.  Thus, these
variables play an important role in the description of thermodynamic evolution.
On the other hand, the field equations determine $\rho$ and $p$ and their
evolution (through the conservation law $T^{\alpha\beta}_{\ \ ;\beta} = 0$),
linking geometrical and physical evolution.
What conditions are necessary and sufficient to guarantee compatibility between
the field equations and the laws of thermodynamics?
According to Coll and Ferrando \cite{colfer},
it is enough to demand the compliance of the integrability conditions of Gibbs
equation. Thus we consider the Gibbs 1-form
\begin{equation}
\Omega = Ts_{,\alpha} {\bf d} x^\alpha
= [(\rho/n)_{,\alpha} + p(1/n)_{,\alpha}
-\mu c_{,\alpha}]{\bf d} x^\alpha
\label{gibbs}
\end{equation}
where a comma denotes partial differentiation and
${\bf d}$ is the exterior derivative. Its
integrability conditions are
\begin{equation}
{\bf d} \Omega = 0
\end{equation}
\begin{equation}
{\bf d}\Omega \wedge \Omega  = 0
\end{equation}
the first of which is a sufficient condition, of no physical significance, as it
involves the existence of a ``heat function" which would imply unphysical
properties to the fluid. The second condition, however, is a necessary and
sufficient condition.
It represents the differential relations that the equations of
state (4 in the case of a binary mixture in 1 phase) must satisfy in order to
contain all the equilibrium information about the system, since not all of them
are independent (only 3 in our case), as one of them can be deduced from the
others via the Gibbs-Duhem equation. Its components are
\begin{equation}
{\bf d}\Omega\wedge \Omega = Z_{tij} {\bf d}t\wedge  {\bf d}x^{i}
\wedge {\bf d} x^j +
Z_{ijk}{\bf d}x^{i} \wedge {\bf d}x^{j} \wedge {\bf d}x^{k}
\end{equation}
\begin{equation}
Z_{tij} = - T\left({1\over n^2} p_{[,t}n_{,i} + \mu_{[,t} c_{,i}\right)
 s_{,j]} = 0
\label{first}
\end{equation}
\begin{equation}
Z_{ijk} = - T\left({1\over n^2} p_{[,i}n_{,j} + \mu_{[,i} c_{,j}\right)
 s_{,k]} = 0
\label{second}
\end{equation}
where square brackets denote antisymmetrization. Using Gibbs
1-form, one can easily see that for the case $c = 1$, {\it i.e.},
for a one component fluid, one obtains conditions equivalent to
those used by Krasi\'nski et al. \cite{ahr,hr1,hr2}.

Then, we say that a given cosmological model with a binary mixture of
perfect fluids as source satisfies the thermodynamic scheme if the integrability
conditions (\ref{first}) and (\ref{second}) are fulfilled and if it evolves
in accordance with the entropy production law (\ref{6}).

\section{Equations of state and evolution}

In this section we exploit the necessary and sufficient integrability condition,
derived in the last section, in order
to find the relations linking the new variables $\mu$ and $c$ with the old ones,
and establish the consequences of the entropy production condition. We do this
for two general families of inhomogeneous solutions of Einstein's equations,
namely the Szekeres and the Stephani solutions.

\subsection{Szekeres solutions}

The source in these models is an irrotational and geodesic perfect fluid,
such that we can find local comoving coordinates in terms of which the metric
has the form \cite{[13]}
\begin{equation}
ds^{2} = dt^{2} - e^{2\alpha} dz^{2} - e^{2\beta} (dx^{2} + dy^{2})
\end{equation}
with  $\alpha=\alpha (t,x,y,z)$, $\beta = \beta(t,x,y,z)$, functions
to be determined from the field equations. In these coordinates the
4-velocity is $u^{\mu} = \delta^{\mu}_{0}$, which implies $\dot{u}^{\mu} = 0$ and so
$p = p(t)$. For their study, we must consider separately
the cases $\beta'= 0$ and $\beta \neq 0$,
where $\beta' = \partial \beta / \partial z$.

In the case $\beta' = 0$, the solution is given by
\begin{equation}
e^{\beta} = \frac{\Phi}{1 + \frac{1}{4}k(x^2 + y^2)}
\end{equation}

\begin{equation}
e^{\alpha} = \lambda + \Phi \Sigma
\end{equation}

\noindent
\noindent
where $\Phi = \Phi(t)$, $k$ is an arbitrary constant, $\lambda =
\lambda(t, z)$, $\Sigma$ is determined by

\begin{equation}
\Sigma = \frac{\frac{1}{2}U(x^2 + y^2) + V_1x + V_2y + 2W }{1 +
\frac{1}{4} k (x^2 + y^2)}
\end{equation}

\noindent
\noindent
with $U=U(z)$, $V_1 = V_1(z)$, $V_2 = V_2(z)$, $W=W(z)$ arbitrary
functions, $\Phi$ is given by

\begin{equation}
\frac{2 \Phi_{,tt}}{\Phi} + \frac{\Phi_{,t}^2}{\Phi^2} + \kappa p +
\frac{k}{\Phi^2} = 0
\end{equation}

\noindent
\noindent
and $\lambda$ satisfies

\begin{equation}
\lambda_{,tt} \Phi + \lambda_{,t}\Phi_{,t} + \lambda \Phi_{,tt} +
\lambda \Phi \kappa p = U + kW
\end{equation}

\noindent
\noindent
Both equations can be solved once the choice $p = p(t)$ has been made.
The matter-energy density is given by

\begin{equation}
\kappa \rho = 2 \left( \frac{\lambda \Phi_{,tt}}{\Phi} - \lambda_{,tt} \right) e^{- \alpha}
+ \frac{3 \Phi_{,t}^{2}}{\Phi^2} + \frac{3 k}{\Phi^2}
\end{equation}

\noindent
\noindent
This family of spacetimes has in general no isometries, but when
$(\lambda/\Phi)_{,t} = 0$ the solution reduces to a FRW model.

\noindent
\noindent
In the case $\beta' \neq 0$ we have

\begin{eqnarray}
e^{\beta} = \Phi e^{\nu} \\
e^{\alpha} = h e^{-\nu}( e^\beta )_{,z}
\end{eqnarray}
\noindent
\noindent
with $\Phi = \Phi(t,z)$, $\nu = \nu(x,y,z)$, $h=h(z)$ and
\begin{equation}
e^{-\nu} = A(x^2 + y^2) + 2 B_1x+ 2B_2y + C
\end{equation}
\noindent
\noindent
where $A = A(z)$, $B_1 = B_1(z)$, $B_2 = B_2(z)$, $C = C(z)$ and
$h(z)$ are arbitrary functions, $\Phi$ is defined by
\begin{equation}
\frac{2 \Phi_{,tt}}{\Phi} +
\frac{\Phi_{,t}^2}{\Phi^2} +
\kappa p
\frac{k}{\Phi ^2} = 0
\end{equation}
\noindent
\noindent
where $p = p(t)$ is an arbitrary function and $k = k(z)$ must satisfy

\begin{equation}
A C - B_1^2 - B_2^2 = \frac{1}{4} ( h^{-2} + k )
\end{equation}

\noindent
\noindent
This family has in general no isometries, but the FRW model results when
$\Phi = zR(t)$ and $k = k_0 z^2$, where $k_0$ is a constant. We now
analyze for both families the conditions under which the
thermodynamic scheme is satisfied.

\noindent \noindent Let us consider condition $Z_{ijk} = 0$ for
this solution, in which $p,_{i} = 0$, and so according to
Eq.(\ref{second}) it reduces to
\begin{equation}
\mu_{[,i}c_{,j} s_{,k]} = 0
\end{equation}
If we interpret this condition as an algebraic equation for the spatial
gradients, $\nabla$, of the variables entering it,
the general solution of this equation can be written as
$\nabla s = a \nabla \mu
 + b \nabla c$, with $a$ and $b$ arbitrary functions depending,
in general, on all coordinates ($t,x^i$). On the other hand,
the spatial components of the Gibbs 1-form (\ref{gibbs})
 for a geodesic fluid imply the relation
\begin{equation}
\nabla s = {1\over T} \nabla \left( {\rho +p \over n}\right) -
{\mu \over T} \nabla  c \ .
\end{equation}
Hence we can identify $a = 1/T$ and $b = -\mu /T$ so that
$\nabla \mu   = \nabla [(\rho + p)/n]$. The integration of this last
equation
\begin{equation}
\mu = {\rho + p \over n}
\label{16}
\end{equation}
yields our first equation of state where we are neglecting an
additive function of time for two reasons: first, in a
two--component system in one phase we have only three degrees of
freedom, which, once fixed, determine all the remaining variables
in such a way that if we consider $\rho$, $p$ and $n$ as
independent, we cannot add to the variable $\mu$ an arbitrary
function of time. Second, a chemical potential of this form can
be interpreted in a natural way. In fact, considering the Gibbs
1-form as given in Eq.(\ref{gibbs}) and the chemical potential
(\ref{16}), we have
\begin{equation}
{\bf d}s = { 1\over nT} {\bf d} \rho - {\mu \over nT} {\bf d} n
- {\mu \over T} {\bf d} c
\end{equation}
{\it i.e.}, the chemical potential of the fluid mixture must determine
both the matter flow due to inhomogeneities in the total particle
number density (with uniform composition) and the matter flow due
to inhomogeneities in the composition (with uniform total particle number density).
Now, if we substitute $\mu$
in the spatial components of the Gibbs 1-form, in terms of
$\rho$ and $p$, which are determined from the field equations,
and of $n$ (remember that particle conservation implies
$n = f/\sqrt{\triangle}$,
where $f(x^{i})$ is an arbitrary function of spatial coordinates and
$\triangle = det(g_{ij})$), then we are in position to recognize the rest
of the equations of state of the mixture
\begin{equation}
c = \ln {c_0\over n} \qquad
T = {T_{0}\over n} \qquad
s = {\mu\over T}
\label{18}
\end{equation}
where $c_{0}$ and $T_{0}$ are functions of time only. Thus, the system
can accommodate, among others, the ideal gas equation of state, which,
as simple as it is, is more realistic than the barotropic one.

Consider now the condition $Z_{tij} = 0$.
For a geodesic fluid it can be written as the equation
\begin{equation}
{{\dot p} \over n^2} \nabla s  \times \nabla n +
{\dot s} \nabla c  \times \nabla \mu +
{\dot \mu} \nabla s  \times \nabla c +
{\dot c} \nabla \mu  \times \nabla s = 0 \
\end{equation}
which by using the components of the Gibbs 1-form (\ref{gibbs}) reduces to
\begin{equation}
{\dot p \over n}  \ \nabla s \ \times \left({1\over n} \nabla n +
\nabla c  \right) = 0
\end{equation}
This condition is trivially satisfied in view of the equation of state for $c$.
Finally, we must demand that the set of thermodynamic variables satisfy
the entropy production condition (\ref{6}). By inserting the corresponding
values given in Eqs.(\ref{16}) and (\ref{18}), it can be shown that
all the thermodynamic variables become functions of time only,
and that the arbitrary functions which define the metric take the values
corresponding to the FRW limit. Consequently, the Szekeres family of solutions
does not satisfy the thermodynamic scheme in general, but only in the
limiting FRW case.

\subsection{Stephani solutions}

These models represent the most general conformally flat solution with an
irrotational perfect fluid as source, admit in general no isometries and generalize
the FRW models. We can find local comoving
 coordinates in terms of which the metric has the form \cite{[13]}
\begin{equation}
ds^{2} = D^{2} dt^{2} - V^{-2} (dx^{2} + dy^{2} + dz^{2})
\label{20}
\end{equation}
\noindent
\noindent
where
\begin{equation}
D = \frac{F V_{,t} }{V}
\end{equation}
\begin{equation}
V = \frac{1}{R} \left\{ 1 + \frac{1}{4} k \left[ (x-x_0)^2 + (y -
y_0)^2 + (z-z_0)^2 \right] \right\}
\end{equation}

\noindent
\noindent
and $F(t)$, $R(t)$, $k(t)$, $x_0(t)$, $y_0(t)$ and $z_0(t)$ are
arbitrary functions of time. The 4-velocity and state variables for
this metric are given by

\begin{eqnarray}
u^{\alpha} = D^{-1} \delta^{\alpha}_0 \\
\rho = 3 C^2  \\
p = -3 C^2 + 2 \frac{VCC_{,t}}{V_{,t}} \\
n = f V^3
\end{eqnarray}

\noindent
\noindent
where $f(x^i)$ is an arbitrary function of spatial coordinates and
$C(t)$ is defined by

\begin{equation}
k = R^2 \left[ C^2 - \frac{1}{F^2}\right]
\end{equation}

\noindent
\noindent
The FRW limit for this solution is obtained when $k$, $x_0$, $y_0$ and
$z_0$ are all constants, or equivalently, when $V_{,t}/V$ is
independent of the spatial coordinates $\{ x, y, z \}$.

To study the thermodynamic behavior of these solutions as
a binary mixture of perfect fluids, we first analyze
the condition $Z_{ijk}=0$. Then, considering that in this case
the thermodynamic
system is the same as in the Szekeres models, we demand that the
equations of state [cf. Eqs.(\ref{16}) and (\ref{18})]

\begin{equation}
\mu = {\rho + p \over n } \qquad s ={\mu \over T}
\label{21}
\end{equation}

\noindent
\noindent
be satisfied. Moreover, introducing Eqs.(\ref{21}) into the Gibbs 1-form and considering
that now $\rho_{,i}=0$, we find that
\begin{equation}
c = \ln {c_0\over ns}
\end{equation}
with $c_{0}$ again a function of time only. On the other hand,
the condition $Z_{tij} = 0$ can be written as
\begin{equation}
{\dot \rho\over n^2} \ \nabla  p \ \times \left({1\over n}  \nabla   n
+ \nabla   c\right) = 0
\label{fst}
\end{equation}
where we have used Eq.(\ref{21}) and the components of the Gibbs 1-form.
Considering now the explicit form of the equations of state for the Stephani solutions,
it can be shown that the integrability condition (\ref{fst})
is satisfied if
\begin{equation}
T = {V\over V_{,t}} {T_{0}\over n}
\label{22}
\end{equation}
where $T_{0}$ is a function of time only.
Furthermore, we have to
demand that the thermodynamic variables so defined satisfy the entropy
production condition (\ref{6}). This can be shown to be equivalent
to demanding that the expression $V_{,t}/V$ depends on time only,
a condition that reduces the Stephani metric to its FRW limit.

\section{Concluding remarks}

We have introduced the set of thermodynamic variables to describe the
most general inhomogeneous cosmological models, considering that the source
is a binary mixture of perfect fluids with, in general, non vanishing
entropy production. This process does not involve forcing isometries on
the metrics. Nevertheless, demanding the compliance of the entropy
production condition reduces the metrics to those of the FRW models.
This condition must be regarded as a consequence
of the local equilibrium hypothesis, so our results indicate certain
incompatibility between thermodynamics and relativity.

We believe that in order to generalize thermodynamics to the
relativistic case, one should first answer
questions about how should one incorporate the second law of
thermodynamics into the thermodynamic scheme and about
the invariance of the thermodynamic variables.
In particular, we note that when irreversible processes
  appear due to inhomogeneities in a system, their main effect is, precisely,
  the vanishing of the inhomogeneities, indicating the need to have a different
  thermodynamic framework, in which non uniform equilibrium states should be
 possible.

\section{Acknowledgements}
  This work was supported  by DGAPA-UNAM grant IN112401,  
CONACyT-Mexico grant 36581-E, and US DOE grant DE-FG03-91ER 40674.
R.D.Z. was supported by a CONACyT Graduate Fellowship.
H.Q. thanks UC MEXUS CONACyT (Sabbatical Fellowship Programm) 
for support.

\end{document}